\documentclass[preprint,number,sort&compress,twocolumn,3p]{elsstyarticle}
\pdfoutput=1
\DeclareMathAlphabet{\scr}{U}{rsfs}{m}{n}
\usepackage{latexsym}
\usepackage{epsfig}
\usepackage[mathscr]{eucal}
\usepackage{amsfonts}
\usepackage{amscd}
\usepackage{amsmath}
\usepackage{array}
\usepackage{amssymb}
\usepackage{colordvi}
\usepackage{enumerate}
\usepackage{graphicx}
\usepackage{booktabs}
\usepackage[footnotesize]{caption}
\usepackage{fancyhdr} 
\usepackage{pdfpages}
\usepackage{slashed}
\usepackage{tabularx}
\usepackage{longtable}
\usepackage{array}
\usepackage[draft]{hyperref}
\usepackage{relsize}
\usepackage{color}
\usepackage{rotating}
\usepackage{cleveref}

\hypersetup{
	colorlinks=true,        
	linkcolor=blue,          
	citecolor=black,        
	filecolor=red,      
	urlcolor=cyan           
}

\definecolor{ourbrown}{RGB}{155,100,15}
\definecolor{ourcyan}{RGB}{20,165,165}
\definecolor{ourpurple}{RGB}{145,0,140}
\definecolor{darkorange}{RGB}{225,100,0}
\definecolor{darkgreen}{RGB}{0,170,0}
\definecolor{darkgray}{RGB}{80,80,80}

\newcommand{\newc}{\newcommand}
\newc{\EW}{electroweak\;}
\newc{\DM}{dark matter\;}
\newc{\SM}{standard model\;}
\newc{\KK}{Kaluza-Klein\;}
\newc{\ff}{fragmentation function\;}
\newc{\be}{\begin{equation}}
\newc{\ee}{\end{equation}}
\newc{\bi}{\begin{itemize}}
\newc{\ei}{\end{itemize}}
\newc{\benu}{\begin{enumerate}}
\newc{\eenu}{\end{enumerate}}
\newc{\bc}{\begin{center}}
\newc{\ec}{\end{center}}
\newc{\bfig}{\begin{figure}}
\newc{\efig}{\end{figure}}
\newc{\neutone}{\tilde{\chi}^0_1}
\newc{\sigmav}{\langle\sigma v \rangle}
\newc{\lamhs}{\lambda_{H\!S}}
\newc{\logJ}{\log (J_{40^\circ}/J_{40^\circ\!,\,\text{nom}})}
\newc{\sigJ}{\sigma_{\log\!J}}
\DeclareMathSymbol{\Gamma}{\mathalpha}{letters}{"00}
\DeclareMathSymbol{\Delta}{\mathalpha}{letters}{"01}
\DeclareMathSymbol{\Theta}{\mathalpha}{letters}{"02}
\DeclareMathSymbol{\Lambda}{\mathalpha}{letters}{"03}
\DeclareMathSymbol{\Xi}{\mathalpha}{letters}{"04}
\DeclareMathSymbol{\Pi}{\mathalpha}{letters}{"05}
\DeclareMathSymbol{\Sigma}{\mathalpha}{letters}{"06}
\DeclareMathSymbol{\Upsilon}{\mathalpha}{letters}{"07}
\DeclareMathSymbol{\Phi}{\mathalpha}{letters}{"08}
\DeclareMathSymbol{\Psi}{\mathalpha}{letters}{"09}
\DeclareMathSymbol{\Omega}{\mathalpha}{letters}{"0A}
\DeclareMathSymbol{\varGamma}{\mathalpha}{operators}{"00}
\DeclareMathSymbol{\varDelta}{\mathalpha}{operators}{"01}
\DeclareMathSymbol{\varTheta}{\mathalpha}{operators}{"02}
\DeclareMathSymbol{\varLambda}{\mathalpha}{operators}{"03}
\DeclareMathSymbol{\varXi}{\mathalpha}{operators}{"04}
\DeclareMathSymbol{\varPi}{\mathalpha}{operators}{"05}
\DeclareMathSymbol{\varSigma}{\mathalpha}{operators}{"06}
\DeclareMathSymbol{\varUpsilon}{\mathalpha}{operators}{"07}
\DeclareMathSymbol{\varPhi}{\mathalpha}{operators}{"08}
\DeclareMathSymbol{\varPsi}{\mathalpha}{operators}{"09}
\DeclareMathSymbol{\varOmega}{\mathalpha}{operators}{"0A}

\renewcommand{\vec}[1]{\boldsymbol{#1}}

\newcommand{\E}{\mathrm{e}}

\newcommand{\hn}{h^0}
\newcommand{\Hn}{{H^0}}
\newcommand{\An}{A^0}
\newcommand{\Hp}{{H^\pm}}

\def\bea{\begin{eqnarray}}
\def\eea{\end{eqnarray}}
\newcommand{\stau}{{\widetilde{\tau}}}

\newcommand{\mstau}{m_{\stau_1}}

\newcommand{\thest}{\theta_{\stau}}
\newcommand{\s}[1]{\widetilde{#1}}
\newcommand{\MEV}{\ensuremath{\,\textnormal{MeV}}}
\newcommand{\GEV}{\ensuremath{\,\textnormal{GeV}}}
\newcommand{\TEV}{\ensuremath{\,\textnormal{TeV}}}
\newcommand{\smo}{\textsc{SModelS}}
\newcommand{\smoversion}{\textsc{SModelS}\,v1.2}

\begin{document}

\title{{Constraining new physics with searches for long-lived particles:\\  
Implementation into SModelS}}

\date{August 15, 2018}
\author[a]{Jan Heisig}
\address[a]{Institute for Theoretical Particle Physics and Cosmology, RWTH Aachen University, 52056 Aachen, Germany}
\ead{heisig@physik.rwth-aachen.de}
\author[b]{Sabine Kraml}
\address[b]{Laboratoire de Physique Subatomique et de Cosmologie, Universit\'e
    Grenoble-Alpes, CNRS/IN2P3,\\ 53 Avenue des Martyrs, F-38026 Grenoble, France}
\ead{sabine.kraml@lpsc.in2p3.fr}
\author[c]{Andre Lessa}
\address[c]{Centro de Ci\^encias Naturais e Humanas, Universidade Federal do ABC, Santo Andr\'e, 09210-580 SP, Brazil}
\ead{andre.lessa@ufabc.edu.br}

\begin{abstract}
We present the implementation of heavy stable charge particle (HSCP) and $R$-hadron 
signatures into \smoversion. We include simplified-model results from the 8 and 13~TeV 
LHC and demonstrate their impact on two new physics scenarios motivated by dark matter:
the inert doublet model and a gravitino dark matter scenario. For the former, we find sensitivity 
up to dark matter masses of 580\,GeV for small mass splittings within the inert doublet, 
while missing energy searches are not able to constrain any significant part of the 
cosmologically preferred parameter space. For the gravitino dark matter scenario, 
we show that both HSCP and $R$-hadron searches provide important limits, 
allowing to constrain the viable range of the reheating temperature.
\end{abstract}

\maketitle

\section{Introduction}\label{sec:intro}

Exploring physics beyond the standard model (BSM) is one of the key scientific goals of the LHC.
Simplified models have turned out to provide useful 
benchmarks for interpreting LHC results and investigating their implications for  
the open questions of today's fundamental physics. The public code 
\smo~\cite{Kraml:2013mwa,Ambrogi:2017neo} provides a very efficient framework for this
reinterpretation by decomposing the signal of an arbitrary new physics model (respecting a ${\cal{Z}}_2$ 
symmetry or a larger symmetry with a ${\cal{Z}}_2$ subgroup) into simplified-model topologies.
This then allows us to apply the simplified-model cross-section upper limits or efficiency maps provided 
by the experimental collaborations (typically as a function of the BSM masses involved) in the context of the new model.%
\footnote{A certain degree of approximation results from 
neglecting properties like the exact production mechanism and the spin of the 
particles in decays, see Refs.~\cite{Edelhauser:2014ena,Edelhauser:2015ksa,Arina:2015uea,Kraml:2016eti} 
for a corresponding discussion.}

So far only BSM searches in final states with missing transverse momentum (MET) could be taken into account within \smo. 
However, it has widely been recognized recently that well-motivated 
BSM theories can provide non-neutral long-lived particles (LLPs) leading to distinct signatures,
which often provide great sensitivity at the LHC~\cite{LLPwhitepaper}.

A method for (re)using LLP simplified-model constraints was developed previously by some of us in \cite{Heisig:2015yla}. 
In this Letter, we now present the implementation of HSCP and $R$-hadron\footnote{For concreteness we label electrically charged but color neutral heavy stable particles as HSCPs. Long-lived colored particles, which can hadronize and form electrically charged bound states are always referred to as $R$-hadrons.} signatures in \smoversion\ and demonstrate its usability to constrain arbitrary BSM scenarios containing such signatures. The models considered as illustrative examples are the inert doublet model (IDM)  and 
the minimal supersymmetric standard model (MSSM) with a gravitino as the ligtest supersymmetric particle (LSP) and a stau next-to-LSP (NLSP). 
In both cases we concentrate on the cosmologically interesting regions of parameter space which satisfy dark matter constraints.

The IDM provides one of the
simplest dark matter models supplementing the standard model by just another SU(2) doublet. 
While MET searches are scarcely sensitive to the cosmologically allowed region of parameter space, 
we show that, for small mass splittings within the inert doublet, a large range of dark matter masses
can be tested with HSCP searches. 

Our second showcase, the gravitino dark matter model, is a cosmologically attractive scenario allowing to
alleviate the gravitino problem and to accommodate large reheating temperatures $T_\text{R}\sim10^9\,$GeV
in the early Universe while respecting bounds from big bang nucleosynthesis (BBN)~\cite{Heisig:2013sva}. 
The complexity of the model reveals a larger number of contributing topologies, including
$R$-hadron signatures relevant for both squarks and gluinos when their decays are 3- or 4-body suppressed.
We show that the LLP results have the potential to be competitive with cosmological constraints and impact the allowed range for the reheating temperature.

With respect to previous versions,  \smoversion\ includes an additional step in the decomposition accounting 
for the probabilities of BSM particles to decay promptly or to appear long-lived. 
Our implementation goes beyond the method presented in \cite{Heisig:2015yla} by incorporating a better treatment of finite lifetimes (for $c\tau$ of the order of the size of detector) and the inclusion of CMS 13 TeV results. 
Concretely, we added the experimental cross-section upper limits for the direct production of HSCPs from \cite{Khachatryan:2015lla,CMS-PAS-EXO-16-036} and $R$-hadrons from \cite{CMS-PAS-EXO-16-036}, and we developed and included recasted efficiency maps for the 13\,TeV analysis~\cite{CMS-PAS-EXO-16-036}. \smoversion\ is publicly available at \href{http://smodels.hephy.at}{http://smodels.hephy.at}.

In the following, Section~\ref{sec:impl} briefly describes the implementation of LLP signatures into \smo. 
The application to the IDM and gravitino dark matter scenarios is presented in Section~\ref{sec:physapp}. 
Section~\ref{sec:summary} contains our conclusions.
Finally, in \ref{app:rec13} and \ref{app:lifetime} 
we provide, respectively, details about the recasting of the HSCP analyses 
and a discussion about the treatment of intermediate lifetimes.

\section{Implementation in \smo}\label{sec:impl}

Given the BSM particle masses, quantum numbers, total decay widths, branching ratios (BRs) and total production cross-sections, 
\smoversion\ performs a decomposition of the input model into a coherent sum of simplified-model topologies. 
Because of the required ${\cal{Z}}_2$ symmetry, the simplified-model topologies have a two-branch structure from the 
production of two BSM states followed by their respective cascade decays. 
We require that each cascade decay terminates
with a long-lived particle (which may lead to MET or a visible LLP signature) while all intermediate decays are required to be prompt. 
By long-lived we mean here and in the following that the particle traverses the whole detector.

\begin{figure*}[h]
\centering
\hspace*{-5mm}\includegraphics[clip, trim={0.8cm 5cm 1.4cm 3.3cm}, width=0.98\textwidth]{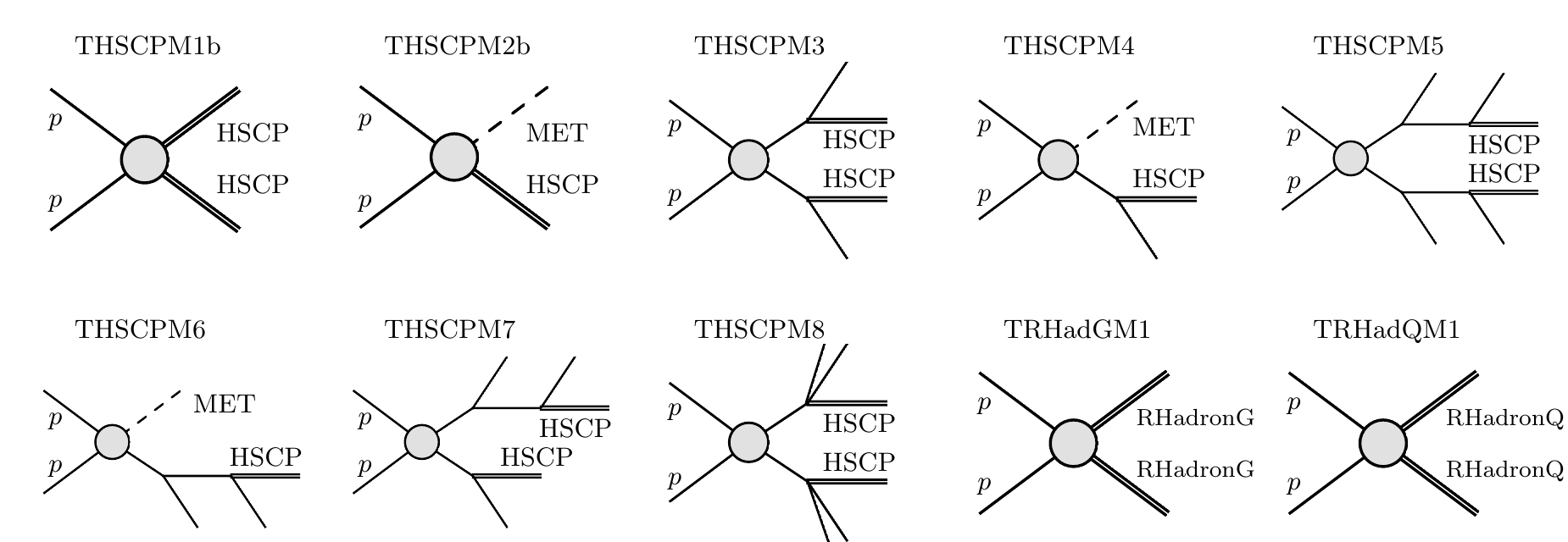}
\caption{HSCP and $R$-hadron simplified model topologies included in the \smoversion\ database.
The double lines represent a lepton-like HSCP, a gluino-like or a squark-like $R$-hadron final state, while dashed lines represent any
cascade decay ending up in a neutral LLP (MET signature). The labels of each topology corresponds to
the naming convention used in the database. Note that in \smoversion\ the
formalism used for mapping the BSM model topologies to the constraints in the database now
includes a list of the BSM final state signatures, e.g.\ (MET, HSCP).}
\label{fig:txnames}
\end{figure*}

At each step in the cascade we compute the 
probability for the BSM particle to decay promptly ($\mathcal{F}_\text{prompt}$) or to decay outside the detector ($\mathcal{F}_\text{long}$), 
which can be relevant for metastable particles with proper lifetimes within few cm to few m~\cite{Heisig:2015yla}.
The respective weight of a topology, $\tilde{\sigma}$, is hence given by: 
\begin{equation}
\label{eq:weights}
\tilde{\sigma} = \sigma_\text{prod}\left(\prod_i
\text{BR}_i \times \mathcal{F}_\text{prompt}^i \right)
\mathcal{F}_\text{long}^X \mathcal{F}_\text{long}^Y \,,
\end{equation}
where $\sigma_\text{prod}$ is the production cross-section of the mother particles and
$X,Y$ are the BSM final states of the two decay chains; the index $i$ runs over all intermediate 
BSM particles. Using the particle proper lifetime ($\tau$), 
$\mathcal{F}_\text{prompt}$ and $\mathcal{F}_\text{long}$ 
can be approximated by:
\begin{equation}
\label{eq:fprompt}
\mathcal{F}_\text{prompt} = 1-\exp\left(-\frac{1}{c\tau}\left\langle\!\frac{\ell_\text{inner}}{\gamma \beta} \!\right\rangle_{\!\!\text{eff}\,}\right)
\end{equation}
and
\begin{equation}
\label{eq:Flong}
\mathcal{F}_\text{long} = \exp\left(-\frac{1}{c\tau}\left\langle\!\frac{\ell_\text{outer}}{\gamma \beta} \!\right\rangle_{\!\!\text{eff}\,}\right)\,,
\end{equation}
where we choose $\langle\ell_\text{inner}/\gamma\beta\rangle_\text{eff}=1\,$mm and $\langle\ell_\text{outer}/\gamma\beta\rangle_\text{eff}=7\,$m.\footnote{This reflects the size of the CMS detector, as we currently have only CMS results included; when including ATLAS results in the future, $\langle\ell_\text{outer}/\gamma\beta\rangle_\text{eff}$ will have to be adjusted accordingly.}
A detailed motivation of the above expressions and values is given in~\ref{app:lifetime}. Note that this choice is conservative, since the signature of particles decaying inside the detector may provide additional sensitivity. However, such cases introduce a dependence on the decay products of the long-lived particle which is left for future work.

Topologies terminating in neutral LLPs (including stable neutral BSM particles) lead to the conventional MET signatures, which constitute the bulk of supersymmetry (SUSY) searches already included in the \smo\ database~\cite{Kraml:2013mwa,Ambrogi:2017neo,Dutta:2018ioj}. 
In order to be able to test also topologies terminating in 
electrically or color-charged BSM states, we have added the CMS searches for lepton-like HSCPs, color-octet (gluino-like) $R$-hadrons and color-triplet (squark-like) $R$-hadrons at 8~TeV~\cite{Chatrchyan:2013oca} and 13~TeV~\cite{CMS-PAS-EXO-16-036} centre-of-mass energies. 

For the 8~TeV  HSCP results, we make use of the recasting provided in Ref.~\cite{Khachatryan:2015lla}, while a dedicated recasting was performed for the 13~TeV results (see~\ref{app:rec13} for details). This allowed us to compute efficiency maps for the eight simplified-model topologies introduced in Ref.~\cite{Heisig:2015yla}, which are included in the \smoversion\ database. 
For the $R$-hadron searches we consider only the direct production topology\footnote{As $R$-hadrons are strongly produced, their production via cascade decays is typically expected to be less important.} and make use of the cross-section upper limits 
from Refs.~\cite{Chatrchyan:2013oca,CMS-PAS-EXO-16-036}
for the cloud hadronization model (assuming a 50\% probability for $\tilde g$-gluon bound state formation).
These limits are also included in the \smoversion\ database.
The relevant topologies and their notation in \smoversion\ are summarized in Fig.~\ref{fig:txnames}.

\section{Physics applications}\label{sec:physapp}

To demonstrate the physics impact, in the following we apply \smoversion\ to two BSM scenarios and derive the 
LHC constraints on their parameter space. As mentioned in the introduction, we consider the IDM as well as the MSSM with a gravitino LSP
and a stau NLSP as illustrative examples.

\subsection{The inert doublet model}\label{sec:IDM}

The IDM is a two-Higgs doublet model with an exact ${\cal{Z}}_2$ symmetry, 
under which all standard model fields (including the Higgs doublet $H$) 
are assumed to be even, while the second scalar doublet $\Phi$ is odd. 
It supplements the standard model Lagrangian by the gauge kinetic terms for $\Phi$ as well as
additional terms in the scalar potential, which now reads
\begin{equation}
\begin{split}
	V = &\;\mu_1^2 |H|^2  + \mu_2^2|\Phi|^2 + \lambda_1 |H|^4+ \lambda_2 |\Phi|^4 \\
	&+ \lambda_3 |H|^2| \Phi|^2
		+ \lambda_4 |H^\dagger\Phi|^2 \\
		&+ \lambda_5/2\,\big[ (H^\dagger\Phi)^2 + \mathrm{h.c.} \big].
\label{Eq:TreePotential}
\end{split}
\end{equation}
After electroweak symmetry breaking the model contains five physical scalar states with masses given by
\begin{equation}
\begin{split}
	&m_{\hn}^2 = \mu_1^2 + 3 \lambda_1 v^2\,,\quad
	m_{\Hn}^2= \mu_2^2 + \lambda_L v^2\,, \\ 
	&m_{\An}^2 = \mu_2^2 + \lambda_S v^2\,,\quad
	m_{\Hp}^2 = \mu_2^2 + \frac{1}{2} \lambda_3 v^2\,, 
\end{split}
\end{equation}
where
\begin{equation}
	\lambda_{L,S} = \frac{1}{2} \left( \lambda_3 + \lambda_4 \pm \lambda_5 \right)\,.
\end{equation}
Imposing $m_{\hn} \simeq 125.09$\,GeV~\cite{pdg}, we are left with five free physical parameters: $m_{\Hn}$, $m_{\An}$, $m_{\Hp}$, $\lambda_L$ and $\lambda_2$.

\begin{figure*}[h]
\centering
\hspace*{-2.5mm}
\includegraphics[width=0.51\textwidth]{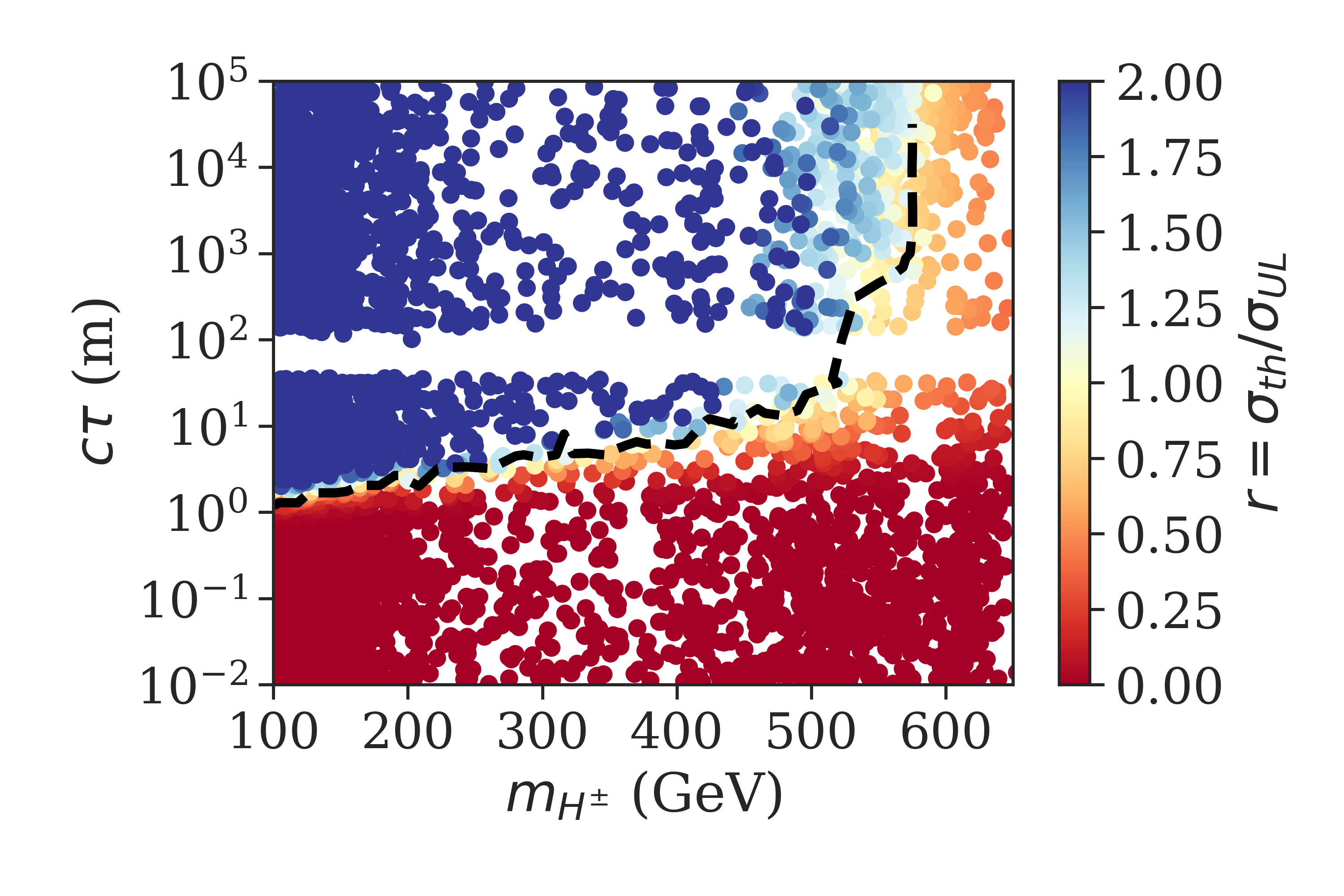}\includegraphics[width=0.51\textwidth]{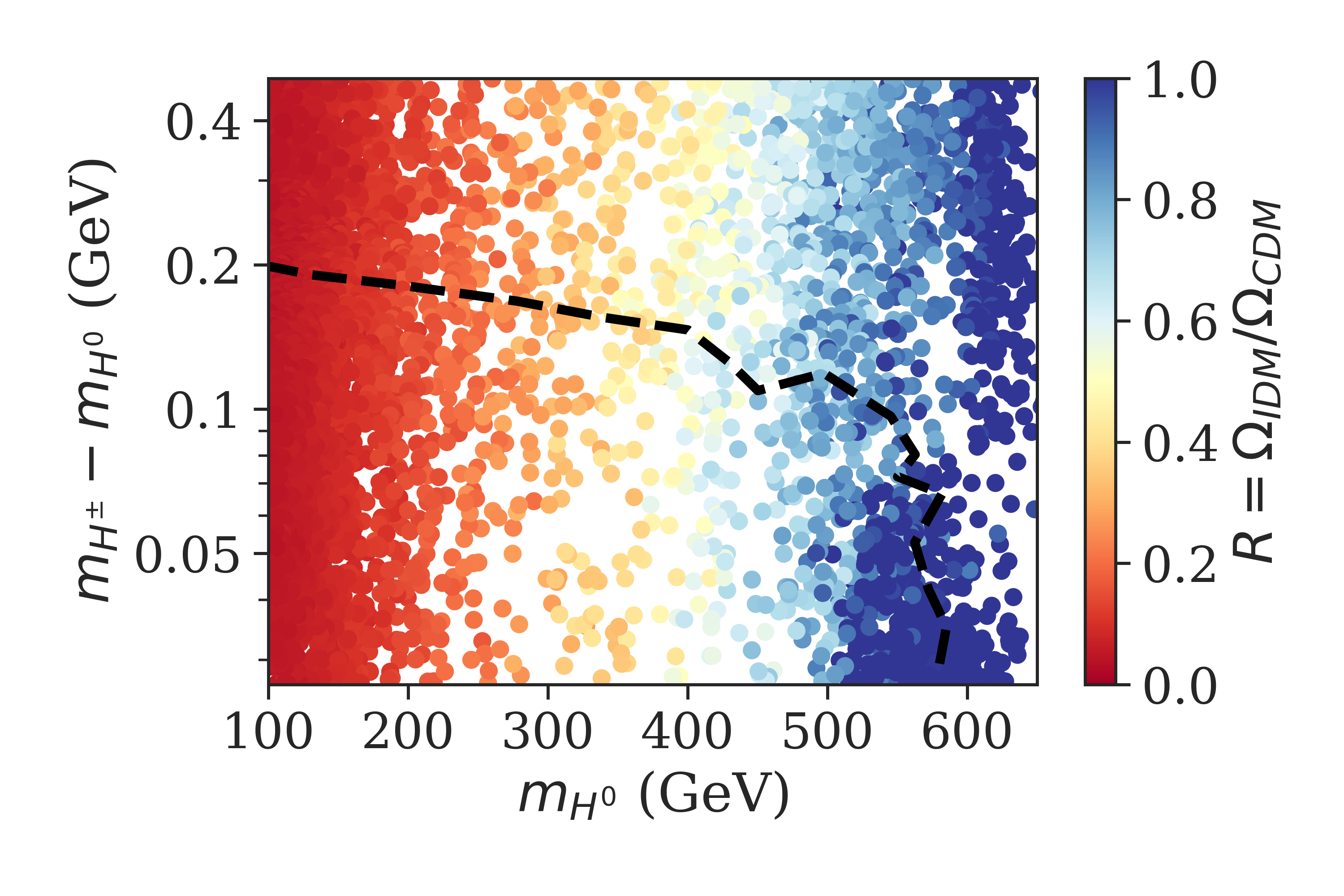}
\vspace*{-8mm}
\caption{Allowed IDM parameter points (imposing all but the LHC constraints) in the $m_{\Hp}$ vs.\ $c\tau$ plane (left panel) and
$m_{\Hp}$ vs.\ $\Delta m$ (right panel). The colour denotes the
LHC signal strength $r$ and the dark matter fraction $R$, respectively. The black dashed curves show the
(interpolated) 95\% CL exclusion contour from the LHC ($r=1$).
}
\label{fig:idmres}
\end{figure*}

Despite its simplicity, the IDM leads to a rich phenomenology and provides a viable dark matter candidate
with observable signatures in direct and indirect detection experiments. For recent accounts 
see, e.g., \cite{Ilnicka:2015jba,Belyaev:2016lok,Eiteneuer:2017hoh}. 
At the LHC the IDM is extremely difficult to observe via MET searches~\cite{Pierce:2007ut,Cao:2007rm,Dolle:2009ft,Miao:2010rg,Gustafsson:2012aj,Belanger:2015kga}. For instance, a reinterpretation of dilepton plus MET searches at 8~TeV~\cite{Aad:2014vma}
provides sensitivity up to $m_{\Hn}\simeq 55$~GeV only~\cite{Belanger:2015kga}. We checked that current 13\,TeV analyses~\cite{Aaboud:2017leg,Aaboud:2018jiw} do not significantly change this picture.
However, in this low-mass region, the $\Hn$ thermal relic density, $\Omega_{\text{IDM}}$, is above
the observed dark matter density, $\Omega_{\text{CDM}}$.
There are in fact three regions where the IDM can account for the entire observed relic density ($55\,\text{GeV}\lesssim m_{\Hn} \le m_{h^0}/2$, $m_{\Hn}\simeq72\,\text{GeV}$ and $m_{\Hn}\gtrsim 500\,\text{GeV}$) and a region where it can account only for a fraction of $\Omega_{\text{CDM}}$  ($72 \lesssim m_{\Hn} \lesssim 500\,\text{GeV}$)~\cite{Goudelis:2013uca,Eiteneuer:2017hoh}.

Considering non-MET signatures the situation is, however, different. As pointed out in~\cite{Belyaev:2016lok} the mass-degenerate region $\Delta m \equiv m_\Hp - m_\Hn \lesssim 1$~GeV is potentially accessible at LHC up to much larger $m_{\Hn}$ in searches for long-lived particles. 
Note also that in the third region mentioned above ($m_{\Hn}\gtrsim 500$~GeV) a rather small mass splitting ($| m_{\Hn} - m_{\An/\Hp} |$ below a few GeV) is required in order to match the relic density constraint~\cite{Hambye:2009pw,Belyaev:2016lok}.

It is therefore interesting to focus on this region with small mass splittings, $\Delta m \lesssim 1$~GeV, and use \smoversion\ to reinterpret the LHC limits from HSCP searches within the IDM model. 
For this purpose we perform a scan over the 5-dimensional parameter space:
\bea
100 \GEV \le & \!m_{H^0} \!&\le1\TEV \nonumber \\
m_{H^0} < & \!m_{A^0} \!& \le1.1\TEV \nonumber \\
10\MEV \le & \!m_{H^\pm} -m_{H^0}\!& \le1\GEV \nonumber \\
-4\pi \le &  \lambda_L  & \le 4\pi  \\
10^{-6} \le &  \lambda_2  & \le 4\pi \nonumber
\eea
imposing $10^{-3} \leq R \equiv \Omega_{\text{IDM}}/\Omega_{\text{CDM}} \leq 1$; 
$\Omega_{\text{IDM}}$ is computed with \textsc{micrOMEGAs}~\cite{Belanger:2008sj,Belanger:2014vza} and we assume a 10\% uncertainty on the prediction. 
In addition, following Ref.~\cite{Eiteneuer:2017hoh}, we take into account constraints from Higgs invisible decays~\cite{Aad:2015pla}, 
electroweak precision observables~\cite{Baak:2014ora,Eriksson:2009ws}, from searches for charginos and neutralinos at LEP-II~\cite{Pierce:2007ut,Lundstrom:2008ai}, 
indirect detection limits from  $\gamma$-ray observations of dwarf spheroidal galaxies~\cite{Fermi-LAT:2016uux}
and theoretical constraints on unitarity, perturbativity and vacuum stability computed with \textsc{2HDMC}~\cite{Eriksson:2009ws}.\footnote{With respect to Ref.~\cite{Eiteneuer:2017hoh}, in this work we update 
    direct detection constraints, additionally imposing the 90\% CL upper limits on the spin-independent
    dark matter-nucleon scattering cross-section recently obtained by Xenon1T~\cite{Aprile:2018dbl}.
}
We use the nested-sampling algorithm \textsc{Multinest}~\cite{Feroz:2008xx,Feroz:2013hea} 
to efficiently explore the parameter space. 
In the following we only consider points allowed within the $2\sigma$ region of the global fit~\cite{Eiteneuer:2017hoh}.

For the allowed parameter space we compute the decay tables and production cross-sections
with \textsc{MadGraph5\_aMC@NLO}~\cite{Alwall:2014hca} and evaluate the LHC constraints with \smoversion. 
For each parameter space point, the constraining power of LHC
searches can be conveniently parametrized by the ratio of the relevant signal cross-section, $\sigma_\text{th}$, to the corresponding analysis upper limit, $\sigma_\text{UL}$: $r \equiv \sigma_\text{th}/\sigma_\text{UL}$ (see Ref.~\cite{Ambrogi:2017neo} for details).
If $r \geq 1$ for at least one analysis, we consider the point as excluded.

The results are shown in Fig.~\ref{fig:idmres}.
In the left panel we display the signal strength $r$
in the $m_\Hp$ vs.\ $c\tau$ plane, while the right panel shows
the dark matter fraction $R$ in the $m_\Hn$ vs.\ $\Delta m$ plane.
Although $r$ does depend on all the model parameters,
it is mostly driven by the mass and lifetime of charged Higgs. We hence show an approximate exclusion curve in
the plot (dark dashed curve). 
In all the parameter space considered, we have verified that
the exclusion is completely dominated by the HSCP searches; even though the MET constraints were also applied, they could not
exclude any of the points. 

In the quasi-stable limit ($c\tau\gtrsim10^3\,$m) HSCP searches exclude $\Hp$ masses up to 580\,GeV.
This limit goes beyond the 13\,TeV LHC limit for direct production of detector-stable staus which reaches 
$m_{\tilde\tau} = 360\,$GeV~\cite{CMS-PAS-EXO-16-036}. The reason for the higher reach is the appearance of 
the additional $W$-mediated production channel $p p \to \Hn\Hp$ as well as (to a lesser extend and depending on $m_{\An}$)
the channels $p p \to \An\Hp,\, \An \An$ with  $\An\to\Hp$.
As an example, for $m_{\Hn} \simeq m_{\An} \simeq m_{\Hn} \simeq 520$~GeV
we have $\sigma(\Hp \Hp) + \sigma(\Hp \Hn) + \sigma(\Hp \An) \simeq 1.14$~fb against $\sigma(\tilde\tau\tilde\tau) \simeq 0.15$~fb~\cite{CMS-PAS-EXO-16-036} for the same stau mass.

At the lower edge of our scan range, for $m_\Hp\simeq m_\Hn\gtrsim100\,$GeV, HSCP searches are able to constrain
decay lengths down to around $c \tau \simeq 2$~m.
Here the significant exponential suppression of ${\cal F}_\text{long}$ is
compensated by the large cross-section. Note that our choice for $\langle\ell_\text{outer}/\gamma\beta\rangle_\text{eff}$
leads to a somewhat conservative exclusion limit in this part of the parameter space (see  \ref{app:lifetime}, left panel of Fig.~\ref{fig:leff}). 

Let us also point out that the calculation of $ c\tau$ involves a certain degree of approximation. In particular, the decays are computed at leading order with \textsc{MadGraph5\_aMC@NLO}, where the decay channels with first generation quarks in the final state are turned off for $\Delta m$ below the pion mass.
This introduces the (artificial) gap seen at $c \tau \sim 100$~m, which would not appear if the proper phase-space including the pion mass were considered.

In the right panel of Fig.~\ref{fig:idmres}, we show the HSCP exclusion curve in the $\Delta m$ vs.\ $m_{\Hn}$ plane.
It excludes a significant part of the parameter space with $\Delta m \lesssim 0.2\,$ GeV. Interestingly, the HSCP searches are starting to exclude part of the region with $R=1$, where the IDM can account for the entire observed relic density ($m_{\Hn} \gtrsim 500$~GeV). 
Note that disappearing track searches have the potential to further extend the reach towards larger $\Delta m$~\cite{Belyaev:2016lok}.

\subsection{Gravitino dark matter scenario}\label{sec:grav}

The gravitino  -- the superpartner of the graviton -- is an attractive dark matter candidate in supersymmetric theories. 
A gravitino ($\tilde G$) LSP is also interesting from the model-building point of view.
Due to strong BBN constraints for gravitinos that decay into lighter sparticles, an unstable gravitino is either required to be extremely heavy or the reheating temperature has to be kept small~\cite{Kohri:2005wn}. While the former limits the viable options for supersymmetry breaking, the latter reduces the options for possible baryogenesis mechanisms. 
In the gravitino DM scenario these problems can be circumvented. 
Thermal gravitino production however still imposes strong constraints on the maximal reheating temperature to avoid overclosure of the universe~\cite{Pradler:2006hh}.

Once the gravitino is the LSP, the lightest sparticle of the MSSM (i.e.~the NLSP) can be any sparticle. However, in order to not reintroduce severe constraints from BBN due to
late decays of the NLSP~\cite{Moroi:1993mb}, certain choices
appear more promising than others. For instance, the stau is an attractive NLSP candidate  providing a large annihilation cross-section, thus resulting in smaller freeze-out abundances. 
Moreover, $\tilde{\tau} \to \tau \tilde G$ decays inject less hadronic (and electromagnetic) energy than the decays of various other NLSP candidates, e.g., $\tilde\chi^0_1 \to Z/\gamma + \tilde G$. 
As a consequence, the impact of the late time stau decays on BBN is reduced. 
At the same time, it also reduces the contribution to the gravitino abundance through NLSP decays. This allows for a larger thermal contribution of gravitino production while not over-closing the Universe. 
Since the thermal contribution is (approximately) proportional
to the reheating temperature, $\Omega^{\text{th}}_{\s G}\propto T_\text{R}$~\cite{Bolz:1998ek,Bolz:2000fu,Pradler:2006hh},
it allows for higher values of $T_\text{R}$, as preferred by classes of models for leptogenesis and
inflation.\footnote{We note that these arguments simply serve to motivate our choice of a stau NLSP, but do not exclude other NLSP candidates.}

We here consider the gravitino DM scenario with a stau NLSP, revisiting the parameter scan performed in~\cite{Heisig:2013rya,Heisig:2013sva} refining and updating the 
constraints from LLP searches at the LHC. 
The scan was performed within the framework of the so-called phenomenological MSSM (pMSSM) with
the additional assumption $ m_{\s q_{1,2}}\equiv m_{\s Q_{1,2}}=  m_{\s u_{1,2}}\!= m_{\s d_{1,2}}$. In this way
a 17-dimensional parameter space was achieved with input parameters and scan ranges given by:\footnote{In this phenomenologically driven parameter scan the spectrum parameters of the third generation sfermions, 
$\mstau, \,m_{\tilde{t}_1}, \,m_{\tilde{b}_1}, \,\thest$ and $ \theta_{\tilde{t}}$, were 
chosen as input parameters in order to obtain an equally good
coverage of small and large mixing scenarios. Tree-level relations were used to translate these 
parameters into soft parameters. In the further analysis only the values recalculated by the spectrum 
generator are used consistently.}
\bea
-10\TEV \le &A_t &\le10\TEV \nonumber \\
-8\TEV \le &A_b,\,A_\tau, \mu& \le8\TEV \nonumber \\
1 \le & \tan\beta & \le 60 \nonumber \\
100\GEV \le &  m_A  & \le 4\TEV \nonumber \\
200\GEV \le & \mstau & \le 2\TEV \nonumber \\
700\GEV \le & m_{\tilde{t}_1}, m_{\tilde{b}_1} & \le 5\TEV \label{eq:scanranges}\\
0 < &  \thest, \theta_{\tilde{t}} & < \pi \nonumber \\
\mstau  \le & \!\!m_{\s L_{1,2}}, m_{\s e_{1,2}} \!\!& \le  4\TEV \nonumber \\
1.2\TEV\le & \!\!m_{\s q_{1,2}}\!\!
& \le  8\TEV \nonumber \\
\mstau  \le &  M_1, M_2 & \le  4\TEV \nonumber \\
1\TEV \le & M_3 & \le  5\TEV \nonumber 
\eea
The particle spectrum was computed with \textsc{SuSpect}~2.41 \cite{Djouadi:2002ze}
and \textsc{FeynHiggs}~2.9.2~\cite{Heinemeyer:1998yj}. 
All points were required to have the lighter stau $\stau_1$ as the NLSP. Moreover, $m_h  \in [123;128]\GEV$~\cite{pdg,Degrassi:2002fi} was imposed.

The decay widths and branching ratios were computed with 
\textsc{SDecay}~\cite{Djouadi:2006bz,Kraml:2007sx} and (in the case of missing dominant decay channels) \textsc{MadGraph5\_aMC@NLO}~\cite{Alwall:2014hca}, while
the freeze-out abundance of staus was computed with \textsc{micrOMEGAs}~\cite{Belanger:2008sj}. 
The analysis considered constraints on the MSSM Higgs sector from
LEP, the Tevatron and the LHC,
EW precision bounds as well as theoretical constrains arising from
charge or color breaking vacua; see \cite{Heisig:2013rya} for details
and references.
In addition, we here impose updated flavor constraints:
$\text{BR}(B\to X_s\gamma) \in [3.0;\;3.64]\times 10^{-4}$
\cite{Amhis:2016xyh} and $\text{BR}(B_s^0\to\mu^+\mu^-) \in
[1.74;\;4.34] \times 10^{-9}$~\cite{Aaij:2017vad}.
Moreover, updated limits on the MSSM Higgs sector are included by applying the conservative constraints on $m_A$ and $\tan\beta$
derived in the $m_h^{\text{mod}+}$ scenario~\cite{Sirunyan:2018zut}. 

Since the gravitino mass can be taken as an additional free parameter, for each point in the pMSSM parameter space satisfying the above requirements, 10 gravitino masses are randomly generated. They are required to lie within an interval where the total $\tilde G$ abundance can match the measured dark matter abundance: 
\begin{equation}
\label{eq:Omcont}
\Omega_{\s G}^{\text{non-th}} h^2+\Omega^{\text{th}}_{\s G}h^2  = \Omega_\text{CDM}h^2\,,
\end{equation}
where $\Omega^{\text{th}}_{\s G}h^2$ and $\Omega_{\s G}^{\text{non-th}} h^2$ are the thermal and non-thermal gravitino relic abundances, respectively.
The latter is generated once the frozen-out $\tilde{\tau}$s decay to $\tilde G$s,
thus enhancing the total gravitino abundance.
The former is determined mostly by the gravitino mass, the gaugino masses and the reheating temperature, $T_\text{R}$~\cite{Bolz:1998ek,Bolz:2000fu,Pradler:2006hh}.
Given a pMSSM point and gravitino mass, first $\Omega_{\s G}^{\text{non-th}} h^2$ is computed and from this the reheating temperature imposing eq.~\eqref{eq:Omcont} with $\Omega_\text{CDM}h^2=0.1189$, taking into account the contributions from the complete spectrum. 
For each of the resulting points in the (17+1)-dimensional parameter space (that by construction fulfills the
relic density constraint), constraints from BBN~\cite{Jedamzik:2007qk,Jedamzik:2006xz} are imposed (see Ref.~\cite{Heisig:2013sva} for further details).

\begin{figure}[t]
    \centering
    \includegraphics[width=0.47\textwidth]{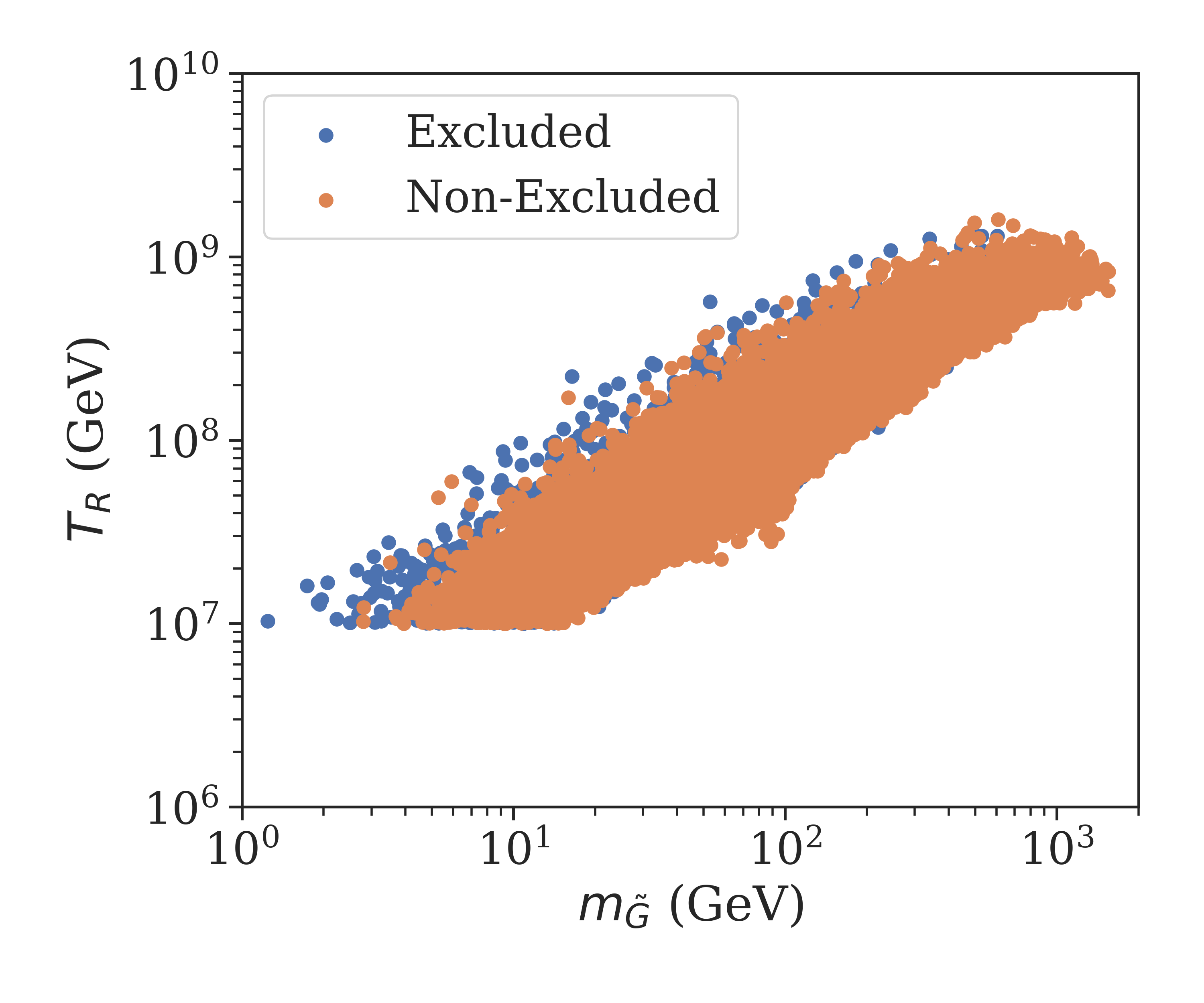}
    \vspace*{-8mm}
    \caption{Effect of the LHC exclusion bounds on the otherwise allowed points in the plane spanned by the gravitino mass and reheating temperature.}
    \label{fig:gavitinores1}
\end{figure}

\begin{figure*}[h]
    \centering
    \setlength{\unitlength}{1\textwidth}
    \begin{picture}(0.96,0.38)
    \put(0.0,-0.00){\includegraphics[clip, trim={0cm 0.1cm 0cm 0cm}, scale=0.5]{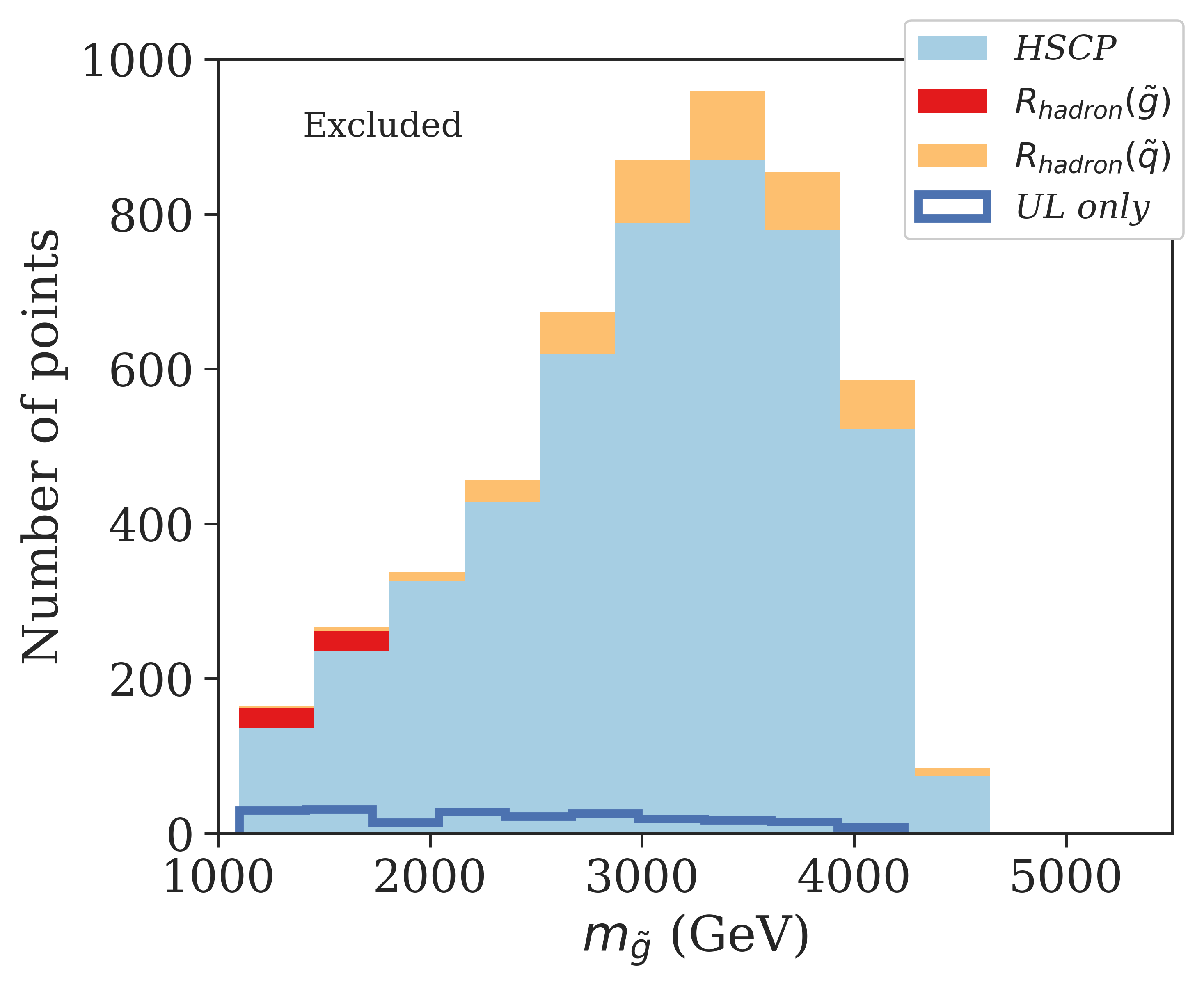}}
    \put(0.5,0.0){\includegraphics[clip, trim={0cm 0.1cm 0cm 0cm}, scale=0.5]{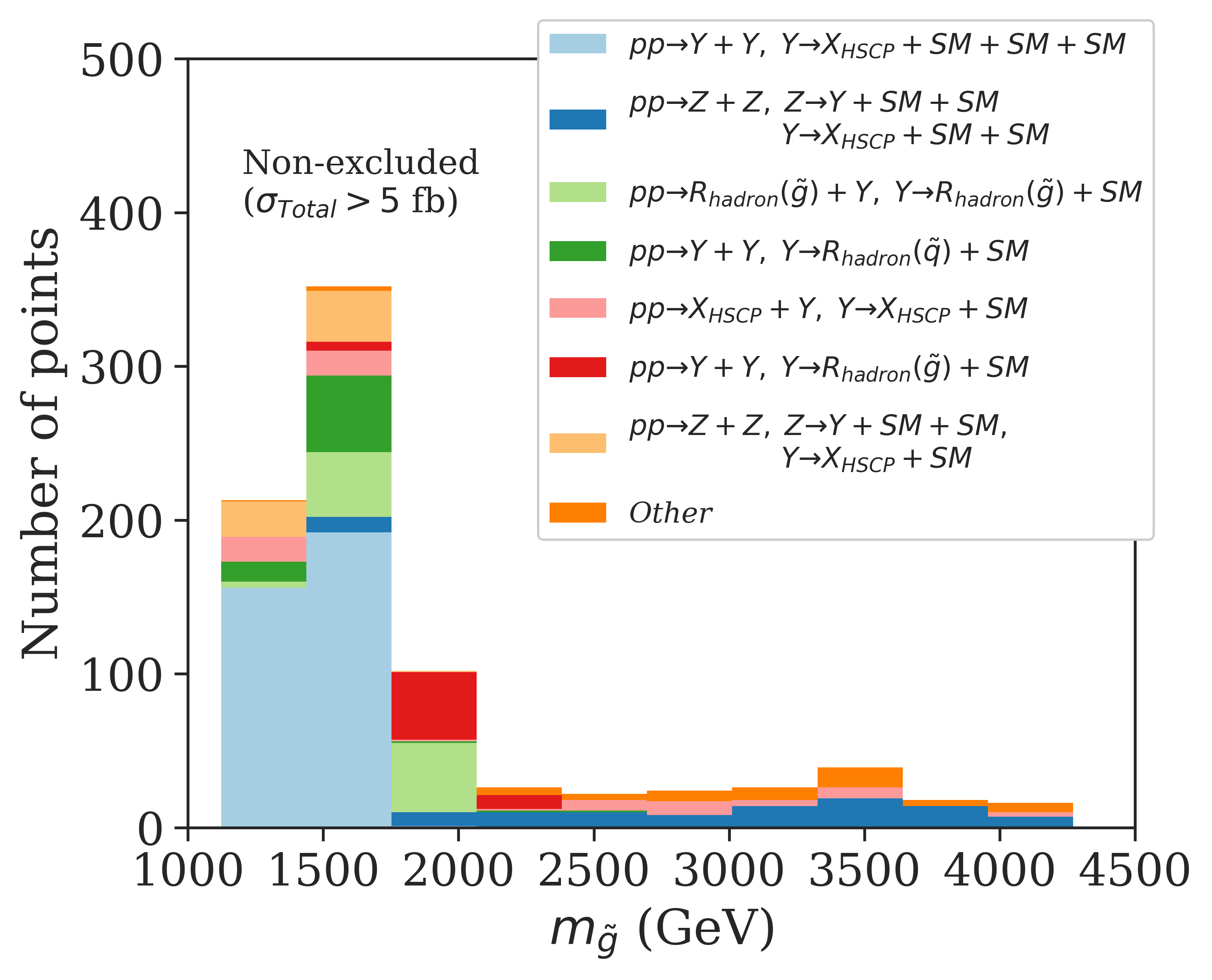}}
    \end{picture}
    \caption{Left panel: Number of excluded points as a function of the gluino mass. The colour indicates the most constraining type of signature. The solid blue histogram displays the number of points excluded when imposing only the CMS limits on the direct pair production of HSCPs and $R$-hadrons.
        Right panel: Number of non-excluded points with a total SUSY production cross-section of more than 5\,fb at 13~TeV. The colour indicates the simplified-model topology with largest weight (cross-section times branching ratio).}
    \label{fig:gavitinores2}
\end{figure*}

After selecting around 26k points satisfying the above constraints,
we use \smoversion\ to decompose each point's LHC signal into all relevant simplified-model topologies, 
taking into account the production and cascade
decays of all sparticles.
The LO production cross-sections are computed with \textsc{Pythia}~8~\cite{Sjostrand:2006za,Sjostrand:2014zea}; 
NLL cross-sections for $\tilde g \tilde g$, $\tilde g \tilde q$ and $\tilde q \tilde q$  are then obtained using \textsc{NLLFast}~\cite{Beenakker:1996ch,Beenakker:1997ut,Kulesza:2008jb,Kulesza:2009kq,Beenakker:2009ha,Beenakker:2010nq,Beenakker:2011fu}.
The results from the 8~TeV and 13~TeV HSCP and $R$-hadron searches implemented in \smoversion\ are then applied in order to constrain the points. Note that, since all the cascade decays terminate (at collider scales) in the stau NLSP or in $R$-hadrons, MET constraints do not apply. 

From all the tested points, $\sim 5$k are excluded and $\sim 21$k are allowed, as shown in Fig.~\ref{fig:gavitinores1}.
For a fixed gravitino mass (below $\sim 200$~GeV), the largest values of $T_{\text{R}}$ are excluded by the LHC constraints. This is due to the fact that the largest reheating temperatures are typically achieved for points with small gluino masses, which in turn contain large production cross-sections at the LHC. As a result these points can be probed by the HSCP and $R$-hadron searches. We see, however, that the largest values of $T_{\text{R}}$ ($\simeq 10^9$~GeV) obtained in the scan are still allowed by the LHC constraints obtained with \smo.

In order to discuss which searches and topologies are relevant for testing the gravitino scenario, we show in Fig.~\ref{fig:gavitinores2} a histogram for the number of excluded points as a function of the gluino mass. 
In the left panel, the number of excluded points is grouped according to which is the most constraining type of signature. The stacked histogram shows that the bulk of the points are excluded by topologies containing HSCP signatures, as expected.
Nonetheless, a significant fraction of points at low $m_{\tilde g}$ are excluded by $R$-hadron constrains for long-lived gluinos. These points typically have heavy squarks, resulting in suppressed 3-body or 4-body gluino decays. In a similar way, points with light squarks and heavy gauginos and higgsinos lead to long-lived squarks which can also be constrained by the $R$-hadron searches, as shown by the orange histogram.

In order to illustrate the constraining power of combining results for multiple simplified-model topologies, we also display (dark blue histogram) the distribution of excluded points obtained using only the CMS limits for pair production of long-lived staus, gluinos and squarks.
As can be seen, the number of excluded points in this case ($\sim200$) is drastically reduced when compared to the one obtained with all the topologies included in \smo.
We point out, however, that the constraining power of \smo\ is still limited by the number of simplified-model results contained in its database. The points from the pMSSM scan performed here
display a large variety of topologies and many of them do not fall within the eight HSCP or the two $R$-hadron topologies included in the 
database.

\smo\ can also conveniently be used to identify the most relevant missing topologies. In the right panel of Fig.~\ref{fig:gavitinores2}, 
we show the non-excluded points with a total SUSY production cross-section (at 13~TeV) larger than 5~fb. Due to their sizeable cross-section, such points have a potential for being excluded by the HSCP or $R$-hadron searches. The stacked histogram shows the distribution of non-excluded points as a function of the gluino mass grouped according to the missing topology with largest weight (cross-section times branching ratio). 
Most of the points have  $m_{\tilde g} < 1.7$~TeV, since this ensures $\sigma(\tilde g \tilde g) \gtrsim 5$~fb.
The almost flat distribution at large $m_{\tilde g}$ corresponds to points with light squarks in the spectrum, thus also resulting in large total cross-sections.

We see that the missing topology which occurs more often in Fig.~\ref{fig:gavitinores2} (light blue histogram) corresponds to pair production of BSM particles, which then go through 4-body decays to the HSCP. This topology is mostly generated by points with very light gluinos, which then decay directly to the $\tilde{\tau}$ through 4-body decays.
Furthermore, we see that topologies with 1-step decays to $R$-hadrons (green and dark red histograms) might also give potentially powerful constraints on this scenario. These topologies often appear in points with light quarks (gluinos) which decay to long-lived gluinos (quarks).

\section{Conclusions}\label{sec:summary} 

Long-lived particles have received increasing attention from
both the experimental and theoretical communities, due to their novel collider signatures 
and possible connection to dark matter.  

In this Letter we discussed how some of the LHC searches for LLPs, interpreted in the context of simplified models, 
can provide a powerful means for constraining interesting dark matter scenarios. 
For this purpose, we presented an implementation of (lepton-like) HSCP and (gluino- and squark-like) $R$-hadron signatures 
into \smo. The \smo\ database was extended by official CMS simplified-model results from the 8 and 13~TeV runs of the LHC,  
as well as additional efficiency maps obtained from a recast of the CMS analyses. 

We then used the new version of \smo, v1.2, to investigate how these searches impact two concrete physics scenarios:
the Inert Doublet Model and a gravitino dark matter model containing long-lived staus. 
While missing-energy searches are not able to constrain any significant
part of the cosmologically interesting parameter space of the IDM, 
we found that, for small mass splittings within the inert doublet of $\Delta m\lesssim 0.2$\,GeV, 
HSCP searches are sensitive to dark matter masses of up to 580\,GeV. 
The gravitino dark matter model, on the other hand, can be constrained by both HSCP and $R$-hadron searches, 
and we showed how \smo\ can be used to derive constraints on the reheating temperature which are 
complementary to cosmological bounds. 

The current version of \smo\ is easily extendible to include additional new simplified-model results from searches for 
detector-stable LLPs from both CMS and ATLAS. The inclusion of other types of LLP signatures, for example 
displaced vertices, is more involved and left for future developments of the code.

\section*{Acknowledgements}

We thank Nishita Desai, Suchita Kulkarni, Ursula Laa and Wolfgang Waltenberger for helpful discussions.
Moreover, very special thanks go to Wolfgang Waltenberger for helping with the public release of the code.

This work is supported by the German Research Foundation DFG through the research unit ``New physics at the LHC''. 
S.K.\ is supported by the IN2P3 project ``Th\'eorie -- LHCiTools'' and the CNRS-FAPESP collaboration grant PRC275431. 
A.L.\ is supported by the Sao Paulo Research Foundation (FAPESP), projects 2015/20570-1 and 2016/50338-6.

\begin{appendix}

\section{Recasting and validation}\label{app:rec13} 

In this appendix we detail the recasting of the 8 and 13~TeV HSCP searches used in \smoversion. We first review the 
recasting for the 8~TeV CMS HSCP analysis presented in~\cite{Khachatryan:2015lla}. The authors of~\cite{Khachatryan:2015lla} provide
signature efficiencies for the off- and online selection criteria, $P_\text{on}(\vec{k})$ and $P_\text{off}(\vec{k})$, respectively, as a function of the generator-level kinematics, $\vec{k}=(\eta,p_\text{T},\beta)$, of isolated\footnote{Details on the imposed isolation criteria can be found in~\cite{Khachatryan:2015lla,Heisig:2015yla}.} HSCP candidates.
The signal efficiency for a given parameter point can be computed from the generated events:
\begin{equation}
\label{eq:signaleff}
({\cal A}\epsilon) = \frac{1}{N}\sum_{i=1}^N {\cal P}^{\,i}_\text{event} 
\end{equation}
where the sum runs over all $N$ events and
\begin{equation}
\label{eq:pevent}
{\cal P}^{\,i}_\text{event} ={\cal P}^{\,i}_\text{on}\times {\cal P}^{\,i}_\text{off}
\end{equation}
with
\begin{equation}
\begin{split}
{\cal P}^{\,i}_\text{on/off} = &\;P_\text{on/off}(\vec{k}_1^i) + P_\text{on/off}(\vec{k}_2^i)\\
& - P_\text{on/off}(\vec{k}_1^i)\times P_\text{on/off}(\vec{k}_2^i)\,.
\end{split}
\end{equation}
For one HSCP candidate in an event the formula holds with $P_\text{on/off}(\vec{k}_2^i)=0$.

\begin{figure*}[t] \centering
\includegraphics[width=0.48\textwidth]{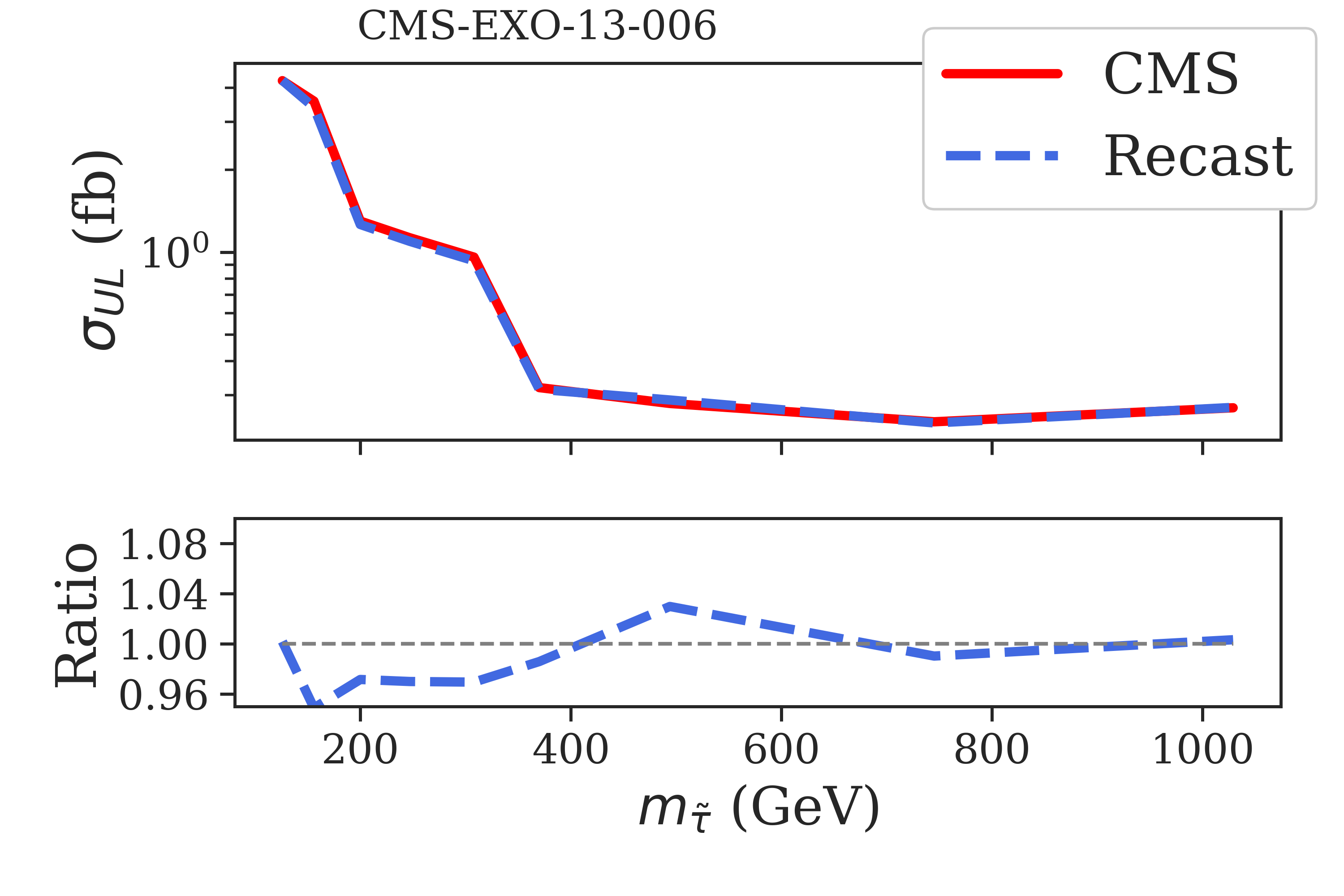}%
\includegraphics[width=0.48\textwidth]{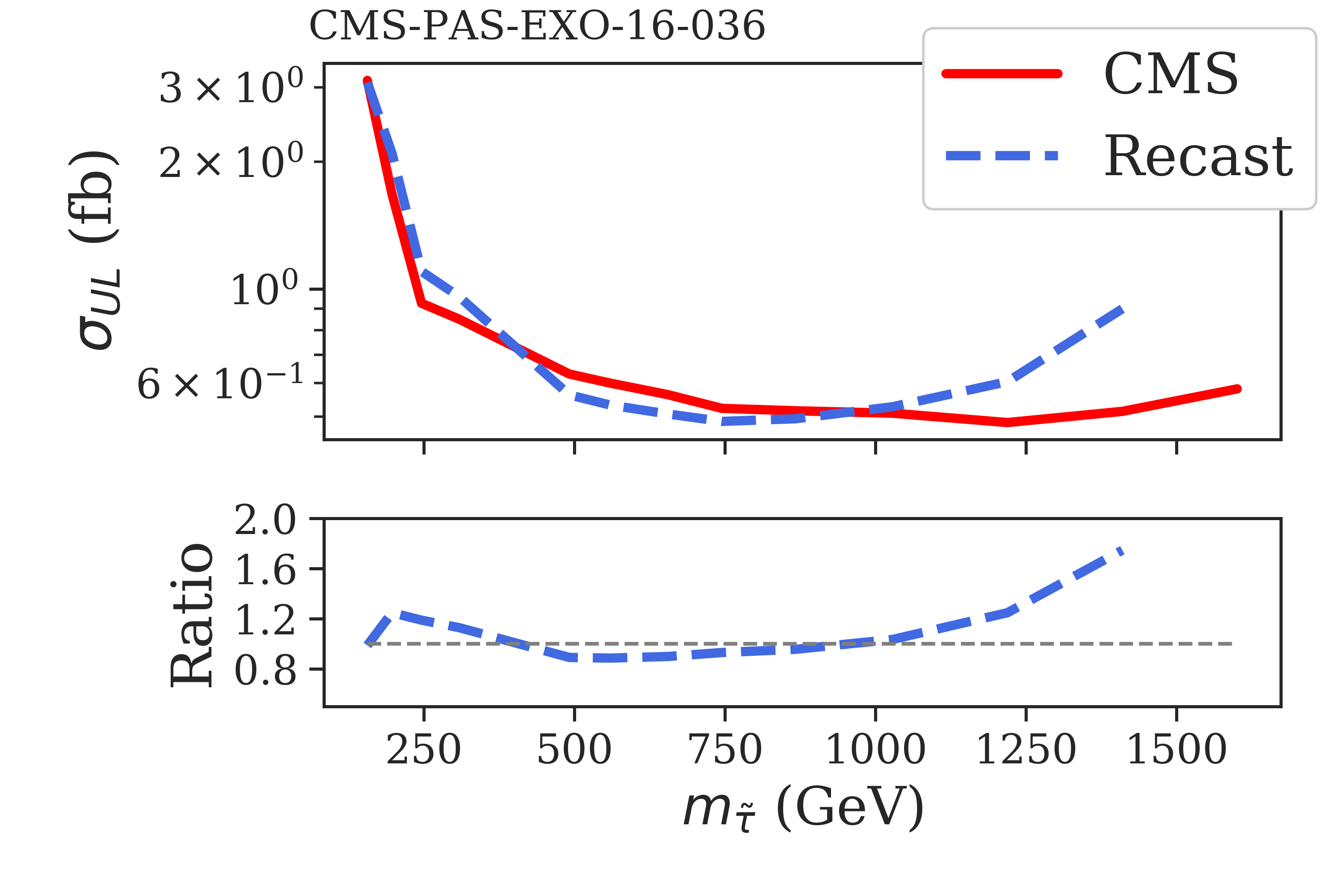}
\vspace*{-4mm}
\caption{Validation of the 8~TeV (left) and 13~TeV (right) CMS analysis for direct production of staus.
The red and blue dashed curves in the upper panels show, respectively, the cross-section upper limits from CMS and from our recast. 
The lower frames show the ratio between our recast and the CMS results.
}
\label{fig:vali}
\end{figure*}

Using the efficiencies computed from \eqref{eq:signaleff} for the direct pair production of staus and the observed and expected number of background events (along with its error) from Ref.~\cite{Khachatryan:2015lla}, 
we obtain upper limits for the stau cross-section as a function of the stau mass. These can then be directly compared to the CMS values presented in Ref.~\cite{Khachatryan:2015lla}. 
The left panel of Fig.~\ref{fig:vali} shows the CMS and our results for the cross-section upper limits (upper frame) as well as their ratio (lower frame). As we can see, the difference is always below 5\% and compatible with Monte Carlo errors. Hence we expect that recasting uncertainties for the efficiencies computed with the above method and included in the \smo\ database should
only be of a few percent.

For the respective 13~TeV analysis~\cite{CMS-PAS-EXO-16-036}, however, such a recast has not yet been provided by the collaboration. Nonetheless, since the trigger and selection criteria of the 8 and 13~TeV analyses are very similar, we 
expect that the signal efficiencies from the 13~TeV search do not
differ drastically from the 8~TeV ones.
The two analyses only differ in a slightly stronger cut on the ionization loss and time-of-flight, which effectively amounts
to a slightly stronger cut on the HSCP velocity in the latter analysis.\footnote{The effect of a slightly stronger cut on $p_\text{T}$~\cite{CMS-PAS-EXO-16-036} is found to be negligible for masses of a few hundred GeV.} 

Ref.~\cite{Brooijmans:2018xbu}
reported an attempt to model the 13~TeV signature efficiencies by multiplying the 8~TeV ones with
a velocity dependent correction function fitted in order to resemble the signal efficiencies reported in~\cite{CMS-PAS-EXO-16-036}.
On top of the slight reduction of the signature efficiencies for high velocities, this study revealed a better performance of the CMS detector in the region of low velocities leading to larger signal efficiencies
for large HSCP masses. This latter feature could, however, not be described by a universal velocity dependent correction function for direct pair production and inclusive production. In order for a proper understanding of the differences between Runs 1 and 2, further information (which is not publicly available) is needed. 

Therefore we choose to follow a conservative approach taking into account the reduction in the efficiency due to the slightly
stronger cuts on the velocity of the HSCP candidate. 
We model this by multiplying $P_\text{off}(\vec{k})$ with a correction function which is assumed to depend only on $\beta$:
\begin{equation}
f_{(a,b)}^\text{corr}(\beta) = \left(1+\E^{a (\beta-b)}\right)^{-1}\le1\,.
\end{equation}
We determine the parameters $a,b$ in a 
global fit to the signal efficiencies for the pair production and inclusive production model reported in Ref.~\cite{CMS-PAS-EXO-16-036}. 
To this end we define the $\chi^2$ function:
\begin{equation}
\label{eq:chi2}
\chi^2 = \sum_m \frac{\left(({\cal A}\epsilon)^m_{(a,b)} - ({\cal A}\epsilon)^m_\text{CMS}\right)^2}{\sigma_{{\cal A}\epsilon}^2}\,,
\end{equation}
where $({\cal A}\epsilon)^m_{(a,b)}$ is the signal efficiency for a mass point $m$ of the considered model
using the signature efficiencies with the correction function $f_{(a,b)}^\text{corr}$ and $({\cal A}\epsilon)^m_\text{CMS}$ is the 
respective signal efficiency reported by CMS in Ref.~\cite{CMS-PAS-EXO-16-036}. The
characteristic size of the uncertainty, $\sigma_{\!{\cal A}\epsilon}$, is (arbitrarily) set to $0.02$, which roughly
reflects the precision of the recasting we aim at.

We minimize the $\chi^2$ using \textsc{Multinest}~\cite{Feroz:2008xx,Feroz:2013hea} 
and obtain the best-fit parameters $a\simeq 500 $ and $b=0.807$. This fit uses all the 12 benchmark points (6 for direct stau production and 6 for inclusive production) considered in Ref.~\cite{CMS-PAS-EXO-16-036} and for which 
signal efficiencies were reported.
We verified that very similar results are obtained when using only a subset of the benchmark points.

Once again we compare the upper limits for the total stau direct production cross-section obtained using our recast procedure and the ones reported by CMS. The comparison is shown in the right panel of Fig.~\ref{fig:vali},
where we see that, despite having a worse agreement than the 8~TeV results,
the 13~TeV upper limits are within 20\% while for large range of stau masses. Only for $m_{\tilde \tau} \gtrsim 1.2$~TeV the 
recasting significantly diverges from the official values. We point out, however, that the results are still conservative due to the above mentioned effects.

\section{Finite lifetimes}\label{app:lifetime}

Although the LLP searches considered here are aimed at detector-stable particles, 
they can also constrain models with intermediate decay length of the order of the detector size, 
where only a certain fraction of particles decay after traversing the entire sensitive detector. In this case the fraction of long-lived particles, ${\cal F}_\mathrm{long}$, may be significantly smaller than one and the resulting signal efficiency becomes sensitive to the value chosen for $L_\text{eff} \equiv \langle\ell_\text{outer}/\gamma\beta\rangle_\text{eff}$ (see eq.~\eqref{eq:Flong}).
Here we discuss in detail what are the expected values for $L_\text{eff}$ and justify our choice, $L_\text{eff}  = 7$~m, 
implemented in \smoversion\ and used in the results presented in Section~\ref{sec:physapp}.

\begin{figure*}[h]
\centering
\setlength{\unitlength}{1\textwidth}
\begin{picture}(1,0.4)
\put(0.0,0.0){\includegraphics[clip, trim={6.15cm 3.5cm 5.5cm 3cm}, width=0.505\textwidth]{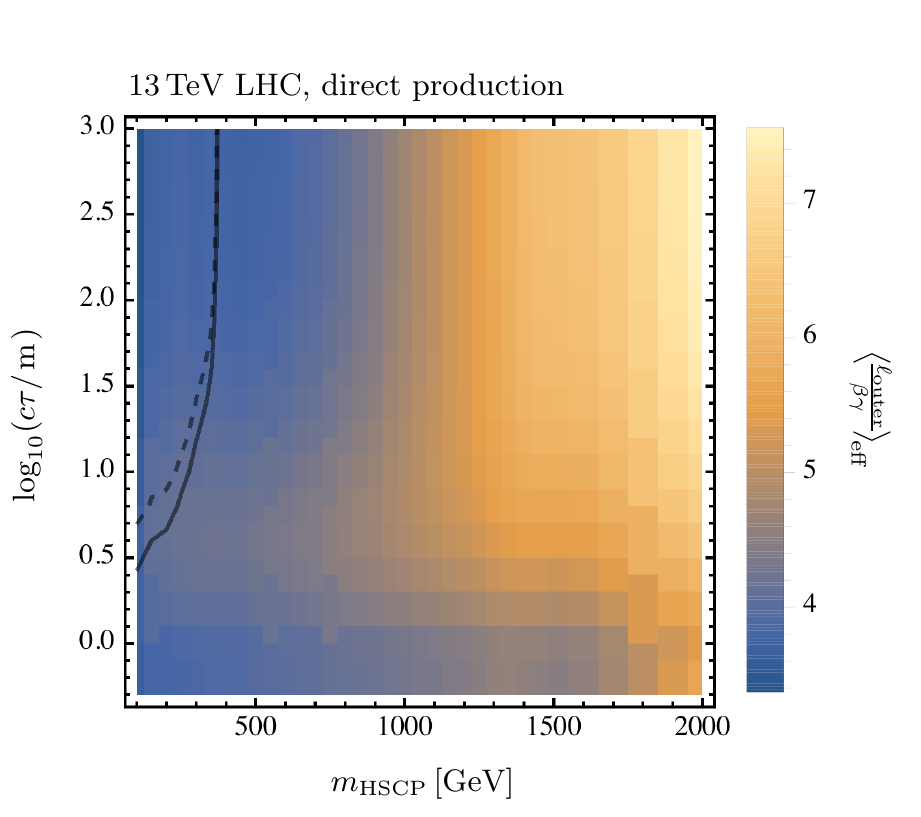}}
\put(0.505,0.0){\includegraphics[clip, trim={6.15cm 3.5cm 5.5cm 3cm}, width=0.505\textwidth]{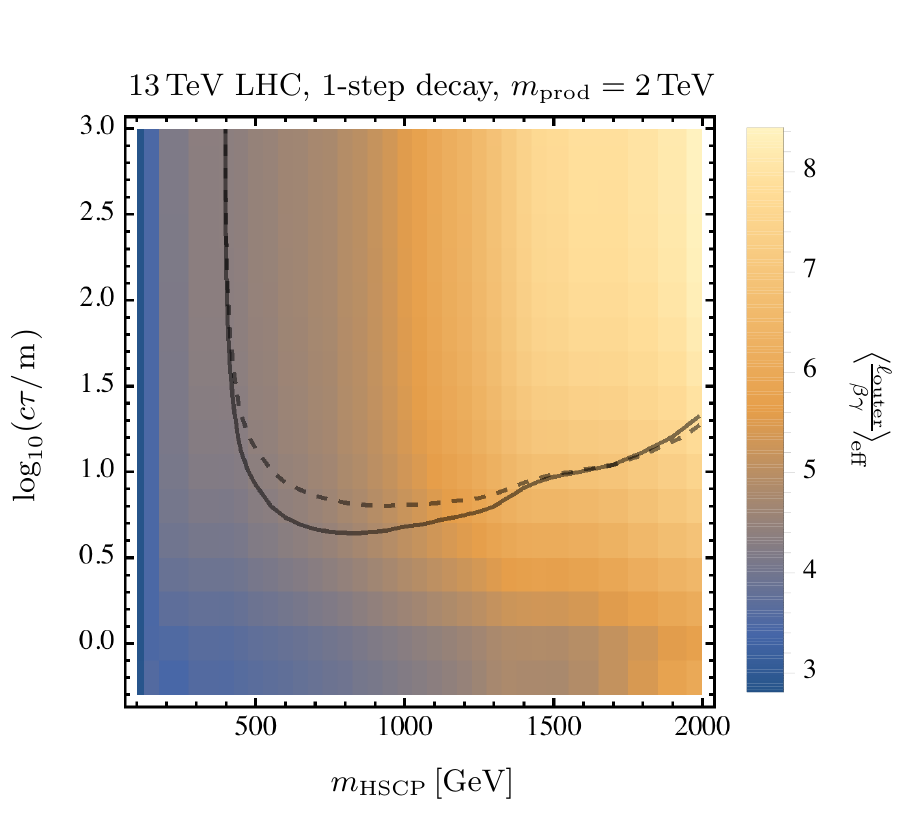}}
\end{picture}
\caption{The effective characteristic travel length, $\langle\ell_\text{outer}/\gamma\beta\rangle_\text{eff}$ in the parameter plane spanned by the HSCP mass, $m_\text{HSCP}$, and its proper decay length, $c\tau$, for direct production (left panel) and for the 1-step decay topology where we choose $m_\text{prod}=2\,$TeV for the mass of the produced mother particle (right panel). The solid and dashed curves denote the 95\% CL exclusion for the event-based computation of $\ell/\gamma\beta$ and for the approximation choosing $\langle\ell_\text{outer}/\gamma\beta\rangle_\text{eff}=7$~m, respectively. For the direct production we choose the cross-section Drell-Yan stau pair production, while the cross-section for 1-step decay corresponds to (degenerate) squark production with $m_{\tilde q}=m_{\tilde g}$.
}
\label{fig:leff}
\end{figure*}

The precise value of ${\cal F}_\mathrm{long}$ (and hence $L_\text{eff}$) 
depends on the input model and experimental analysis and
requires a full Monte Carlo simulation for each model point in order to determine the boost distribution of the LLPs\@. However, since \smo\ aims for a fast (although approximate)
computation of LHC constraints for a large variety of BSM models, our
goal is to determine an average value for $L_\text{eff}$ which can approximately reproduce the correct value of ${\cal F}_\mathrm{long}$ obtained from a full simulation.
Before we can justify this approximation, we must first discuss how to obtain  ${\cal F}_\mathrm{long}$ from the full simulation for a given input model.

We first define the probability $F_\text{long}$ for a (metastable) particle with momentum $\vec{k}$ to decay outside the detector in a given event:
\begin{equation}
\label{eq:flongevt}
F_\text{long}(\vec{k}) = \exp\left(-\frac{\ell_\text{outer}(|\eta|)}{\gamma \beta }\frac{1}{c\tau}\right)\,.
\end{equation}
Here $\gamma = (1-\beta^2)^{-1/2}$ and $\ell_\text{outer}(|\eta|)$ is the travel length
through the CMS detector, which we approximate by considering a cylindrical volume with radius of 7.4~m and length of 10.8~m.
Using now the off- and online efficiencies ($P_{\text{on}}$ and $P_{\text{off}}$) discussed in \ref{app:rec13}, we can
extend the signal efficiency calculation from
eq.~\eqref{eq:signaleff} to the case of finite lifetimes using:
\begin{equation}
\begin{split}
\label{eq:Pevfull}
{\cal P}^{\,i}_\text{event} = &\;
F_\mathrm{long}(\vec{k}_1^i)  P_{\text{on}}(\vec{k}_1^i) P_{\text{off}}(\vec{k}_1^i) \big(1- F_\mathrm{long}(\vec{k}_2^i)\big)\\
&\!\!\!+F_\mathrm{long}(\vec{k}_2^i)  P_{\text{on}}(\vec{k}_2^i) P_{\text{off}}(\vec{k}_2^i) \big(1- F_\mathrm{long}(\vec{k}_1^i)\big)\\
&\!\!\!+ F_\mathrm{long}(\vec{k}_1^i) F_\mathrm{long}(\vec{k}_2^i) {\cal P}^{\,i}_\text{on} {\cal P}^{\,i}_\text{off},
\end{split}
\end{equation}
where $F_\text{long}(\vec{k}_j^i)$ is the decay probability from eq.~\eqref{eq:flongevt} for the $j$-th particle in the $i$-th event.

Using eqs.~\eqref{eq:Pevfull} and \eqref{eq:signaleff} we can then compute the total signal efficiency for a given input model taking into account the correct finite lifetime suppression factor. This efficiency can then be compared to the one computed with \smo\ using ${\cal F}_\mathrm{long}$
to extract the precise value for $L_\text{eff}$ given a specific input model.

In Fig.~\ref{fig:leff} we consider the direct production of staus (left panel) and direct production of squarks followed by a 1-step decay into the HSCP at the 13~TeV LHC.
In both cases we vary the HSCP mass and lifetime, with $m_{\tilde g} = m_{\tilde q} = 2$~TeV in the second case.
The colour of each point in the plane
shows the correct value for $L_{\rm eff} = \langle\ell_\text{outer}/\gamma\beta\rangle_\text{eff}$ 
which should be used in \smo\ in order for the signal efficiencies to exactly match the full simulation values.

As we can see, $L_{\rm eff}$ does not vary significantly, spanning values of about 4--8\,m. Therefore, in order to remain conservative 
and at the same time avoid (significantly) underestimating the signal efficiency, we choose $L_{\rm eff} = 7$\,m (or $\ell_\text{outer} \simeq 10$~m and $\gamma\beta \simeq1.43$). 
Figure~\ref{fig:leff} compares the exclusion curves obtained with \smoversion\ using this approximation (dashed lines) to the ones obtained using the full simulation (solid lines).
Although the \smo\ curves are conservative, we see that they agree quite well with the full simulation in most of the parameter space. 
We have checked that this also holds for other masses and topologies,
though for the sake of conciseness do not show these results. 
We therefore conclude that using a fixed $L_{\rm eff} = 7$~m value is a valid approximation.\\

\end{appendix}

\addcontentsline{toc}{section}{References}
\bibliographystyle{utphys.bst}
\bibliography{SModelS_HSCP_ref}

\end{document}